\documentclass[12pt,epsfig]{article}
\usepackage{graphicx}
\usepackage{eufrak}
\usepackage[mathscr]{eucal}
\usepackage{latexsym}
\usepackage{epsfig}

\def\be{\begin{equation}}
\def\ee{\end{equation}}
\def\ba{\begin{eqnarray}}
\def\ea{\end{eqnarray}}
\def\kk{\mathfrak{K}}

\begin{document}
\baselineskip=15.5pt
\pagestyle{plain}

\rightline{hep-th/0008102}

\rightline{HUTP-00/A033}

\begin{center}

\vskip 1.7 cm

{\LARGE {\bf Comments on Brane World Cosmology}}\\


\vskip 2.5cm
Luis Anchordoqui$^{a,}$\footnote{doqui@hepmail.physics.neu.edu}
and Kasper Olsen$^{b,}$\footnote{kolsen@feynman.harvard.edu}

\medskip
\medskip
${}^a${\it Department of Physics, Northeastern University\\
Boston, MA 02115, USA}

\medskip
${}^b$
{\it Department of Physics, Harvard University\\
Cambridge, MA  02138, USA}

\vspace{2cm}

\end{center}

\noindent
In this paper we consider some constraints on brane-world cosmologies. 
In the first part we analyze different behaviors for the expansion of our 
universe by imposing constraints on the speed of sound.  
In the second part, we study the nature of matter on the 
brane world by means of the 
well-known energy conditions. We find that the strong energy condition must
be completely violated at late stages of the universe.


\newpage

\tableofcontents

\section{Introduction}

Recently there has been a lot of interest in the old idea 
\cite{KK} that spacetime has more than four dimensions 
\cite{renacimiento}. The most attractive scenario 
along these lines is, perhaps,  the so-called 
``Randall Sundrum (RS) brane world'' \cite{lisa}. 
Within this framework the 
background metric is not flat along the extra coordinate; rather it is a 
slice of Anti-de Sitter ($AdS$) space.  
Of course, in this setup, everything is confined to live on the brane except 
for gravity itself, without conflict with observations. 
Generalizations of the RS model including 
embeddings into supergravity or string theory \cite{rs+}, 
as well as a number of interesting phenomenological issues \cite{pheno}, 
have sparked a flurry 
of activity and several groups have begun to search for 
possible experimental 
signatures of these kinds of models \cite{rs-exp}. In addition, 
since the isometry group of the bulk continuum coincides with the conformal 
group of the brane, the 
Maldacena conjecture could be exploited \cite{malda}. Roughly speaking, 
the $AdS/CFT$ correspondence may be able to set uniquely the boundary 
conditions for the fields on the edge. 
(We will briefly touch upon this issue in the next section).

On a different track, the picture of a brane world evolving in a larger 
spacetime gives an interesting new perspective on early universe 
cosmology \cite{cosmology,radion}. In particular, attention has 
been devoted to the question of the 
quantum creation of the world. A series of rather 
recent papers advance the idea 
that a brane bubble can nucleate spontaneously 
together with its $AdS$ bulk \cite{gs,hhr,graftongroup,ch}. 
However, it is not clear yet 
whether this system could evolve towards a configuration of thermal 
equilibrium consistent with experimental data \cite{te}. 
In this paper we elaborate on this question. 
In section 2 we start by considering the very early stages of the brane world
(i.e. at temperatures $T\sim 10^{27}K$). Assuming 
a general equation of state for the matter on the brane, we are able 
to constrain the evolution of the system by using standard procedures of 
shell stability against radial perturbations \cite{visser-book}. Strictly 
speaking, we show that tuning the speed of sound to the standard range, 
while (almost) consistent 
with a de Sitter world, may lead to unusual regimes for $AdS$ domain walls.     Bounds from the Weyl anomaly induced by the $CFT$ that lives on the brane 
are also discussed. Afterwards, in section 3 we turn to much later ages of
the brane world (at
temperatures $T\leq 60K$). We study possible constraints from the energy
conditions. In particular, we find 
that the matter threading the brane world
is not consistent with the strong energy condition unless the present value
of the Hubble constant is $H_0 \approx 30$ km s$^{-1}$ Mpc$^{-1}$ 
(recent measurements find  $H_0 = 67\pm 10$ km s$^{-1}$ Mpc$^{-1}$ 
\cite{tt}). We close in section 4 with a brief discussion.

\section{The Early Brane World}

\subsection{The Very Early Brane World}

According to the Hartle--Hawking ``no boundary'' proposal, the 
quantum state of the universe is given
by an Euclidean path integral over compact metrics 
\cite{HH}. This picture can be adapted to the RS 
scenario by surgical grafting two Euclidean balls 
of $AdS_5$  \cite{gs,hhr}. The 
evolution of the brane after creation is given by the analytical continuation
of $S^5$ to real time, i.e., a de Sitter hyperboloid embedded in Lorentzian 
$AdS_5$ space. This brane world created from ``nothing'' is 
completely analogous to the four--dimensional de Sitter instanton, except that 
here the inside of the wall is filled with $AdS$ bulk.
Following \cite{hhr,graftongroup,gubser} one can use the
$AdS/CFT$ correspondence to explain the behavior of the inflating instanton. 
In particular, $CFT$s generally exhibit a conformal anomaly when coupled to 
gravity \cite{duff}. In the above 
setup, this anomaly is the ``carrier'' of the effective cosmological 
constant on the brane.  

Formally, the above scenario can be generalized to $d$-dimensional de 
Sitter spaces which bound $d+1$ dimensional $AdS$ spaces. The $CFT$, however, 
is not straightforward to obtain. The classical  
action describing the above setup is given by,
\begin{equation}
S = {L_p^{(3-d)} \over 16 \pi}\int_\Omega d^{d+1}x\sqrt{g}
\left(R+ {d\,(d-1)\over \ell^2}\right) + {L_p^{(3-d)} \over 8 \pi}
\int_{\partial \Omega}
d^dx  \sqrt{\gamma} \,\, \kk  + T \int_{\partial \Omega} d^d x
 \sqrt{\gamma}, 
\label{ogete}
\end{equation}
where $\kk$ stands for the trace of the extrinsic curvature of the
boundary, $\gamma$ is the induced metric on the brane, and $T$ is
the brane tension.\footnote{In our convention, the extrinsic
curvature is defined as $\kk_{\Xi\Upsilon} = 1/2 (\nabla_\Xi\hat{n}_\Upsilon
+ \nabla_\Upsilon \hat{n}_\Xi)$, where $\hat{n}^\Upsilon$ is the outward
pointing normal vector to the boundary $\partial\Omega$. Capital 
Greek subscripts run from 0 to $d$ and refer to the entire (d+1) 
dimensional  spacetime, capital Latin subscripts run from 0 to $(d-1)$ 
and will be used to refer to the brane sub-spacetime, lower Latin 
subscripts run from 1 to $d-1$ and refer to constant $t$ slices on
the brane. Throughout the paper we adopt
geometrodynamic units so that $G\equiv1$, $c\equiv1$ and $\hbar
\equiv L_p^2 \equiv M_p^2$, where $L_p$ and $M_p$ are the Planck
length and Planck mass, respectively.} 
(The following discussion will mostly refer to $d=4$, but 
most of the equations will be written for arbitrary $d$).
The first term is the usual
Einstein-Hilbert (EH) action with a negative cosmological constant
($\Lambda=-{1 / \ell^2}$). The second term is the Gibbons-Hawking
(GH) boundary term, necessary for a well-defined variational problem
\cite{gibbons-hawking}. The third term corresponds to a constant
``vacuum energy'', {\it i.e.} a cosmological term on the boundary.

\begin{figure}[t]
\epsfxsize=3.5in
\bigskip
\centerline{\epsffile{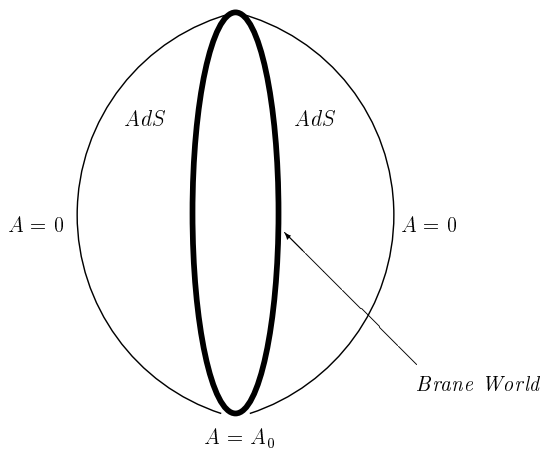}}
\caption{\baselineskip=12pt {\it De Sitter instanton. The brane is
a sphere bounding to $AdS$ Euclidean spaces. (The figure was adapted from
\cite{hms}).}}
\bigskip
\label{instanton}
\end{figure}

Before proceeding further, we note that the following coordinate
transformations: 
\begin{equation}
\frac{d\alpha}{d\phi}=\ell\,\tanh(\eta/\ell)\,\sin\chi\,\sin\beta\,\sin\theta 
\, \,\, ,\,\,\,\,\,\,
\frac{dy}{d\eta}=\left(1+\frac{y^2}{\ell^2}\right)^{1/2},
\end{equation}
give a diffeomorphism between the Euclidean $AdS_5$-metrics
\begin{equation}
ds^2=d\eta^2+\ell^2\sinh^2(\eta/\ell^2)[d\chi^2+\sin^2\chi d\Omega_3^2],
\label{euclidmetric}
\end{equation}
and 
\begin{equation}
ds^2 = \left(1+\frac{y^2}{\ell^2}\right) d\alpha^2 +
\left(1+\frac{y^2}{\ell^2}\right)^{-1} dy^2 + y^2 d\Omega_3^2.
\label{euclidinstanton}
\end{equation}
Here $d\Omega_3^2=d\beta^2+\sin^2\beta(d\theta^2+\sin^2\theta d\phi^2)$ 
is the volume element of the three-sphere (see Fig. \ref{instanton}). 
Thus, one can glue two of these balls along the four sphere 
boundaries, and then  analytically continue to Lorentzian signature 
to obtain a four-dimensional de Sitter brane embedded in Lorentzian $AdS_5$ 
space.
For simplicity we referred the above formulae to Euclidean $AdS_5$ balls, 
but it can be trivially generalized to $AdS_{d+1}$. 

With this in mind, we consider a brane that bounds two regions of 
Lorentzian $AdS$ spaces, which are conveniently described in the static 
chart as 
\begin{equation}
ds^2 = -\left(k+\frac{y^2}{\ell^2}\right) dt^2 +
\left(k+\frac{y^2}{\ell^2}\right)^{-1} dy^2 + y^2 d\Sigma_k^2,
\label{metric}
\end{equation}
where  $d\Sigma_k^2$ is the corresponding metric
on a $(d-1)$ dimensional space of constant curvature with metric 
$\bar{g}_{mn}$, and  Ricci tensor $\bar{R}_{mn} = k (d-2)\bar{g}_{mn}$
with  $k \in \{-1,0,1\}$ corresponding respectively to hyperbolic, flat and
spherical geometry. The case of particular interest 
here is the case $k=1$,
for which the conformal anomaly of the $CFT$ increases the effective 
tension of the domain wall, yielding an everlasting 
inflationary universe \cite{graftongroup}. 
Henceforth, $k$ shall merely be carried along as an arbitrary 
constant that details alternative symmetries for the brane. 
It should be noticed, that when the scale factor is large 
enough the spherical universe would be
practically indistinguishable from a spatially flat universe ($k=0$).

Applying the thin-shell formalism \cite{israel} 
the field equation reads,
\begin{equation}
T g_{_{\Xi \Upsilon}} \delta^{^\Xi}_A \, \delta^{^\Upsilon}_B =
 \frac{L_p^{3-d}}{4\,\pi} [\kk_{AB} - {\rm tr} (\kk) g_{_{\Xi \Upsilon}}
\delta^{^{\Xi}}_A \, \delta^{^\Upsilon}_B].
\end{equation}
The above equation implies that the case $k=1$ corresponds to an inflating 
universe which is forever expanding (a comprehensive analysis of a domain 
wall that inflates, either moving through the bulk or with the bulk 
inflating too, was first discussed by Chamblin--Reall \cite{cosmology}).

\subsection{Matter--driven Expansion}

Inflation will only be useful if it comes to an end. 
In the spirit of \cite{staro} we will 
assume that the world is created with the 
matter fields in their ground state, and when it starts falling under the action 
of the higher dimensional space the matter fields become excited. 
If this is the case, the subsequent expansion results from matter 
on the brane. We take the energy momentum tensor to be 
\begin{equation}
\tilde{T}_{AB} = - T_c \gamma_{AB} + 
\tilde{\rho} \,u_A u_B + \tilde{p} \,(\gamma_{AB} 
+ u_A u_B),
\end{equation}
corresponding to matter with energy density $\tilde{\rho}$ and pressure 
$\tilde{p}$. Here, $u_A$ stands for the  velocity of a piece of stress-energy 
in the co--moving system ($u_A u^A = -1$), tuning the 
vacuum energy to be $T_c=(d-1)/(4\pi\ell L_p^{d-3})$.
From now on, to simplify notation, we denote by 
$\tilde{T}_0^0 \equiv - \rho(A)$, $\tilde{T}_m^m \equiv p(A)$.

The system can be decomposed into falling 
shells (which do not interact with each other or with the environment 
that generates the metric), with trajectories described by the scale 
factor $A(\tau)$. In other words, while the brane-world is sweeping 
through the $(d+1)$ dimensional bulk, the change in the internal energy 
is compensated by the work done by the internal forces, 
\begin{equation}
\frac{d}{d\tau}\, \rho {\cal S} +
p \,\frac{d}{d\tau}\, {\cal S} = 0,
\end{equation}
where, 
\begin{equation}
{\cal S} = \frac{4 \pi A^2}{(d-1) \,L_p^{3-d}}.
\end{equation}
It is straightforward, using Eq. (6) and definitions above, to check that 
\begin{equation}
\rho =  \frac{L_p^{(3-d)}}{4\pi} \frac{(d-1)}{A}\,\left(k+\frac{A^2}{\ell^2} +
\dot{A}^2 \right)^{1/2},
\label{m1}
\end{equation}
and
\begin{equation}
p = - \frac{L_p^{(3-d)}}{4\pi} \left\{
\frac{(d-2)}{A}\left(k+\frac{A^2}{\ell^2} +  \dot{A}^2 \right)^{1/2} +
\frac{\ddot{A} + A/\ell^2}{\sqrt{k +\dot{A}^2 + A^2/\ell^2}} \right\},
\label{m2}
\end{equation}
satisfy the required energy conservation (dots denote derivatives 
with respect to $\tau$). It should also be noted that the jump in the 
second fundamental form selects the positive value of the square-root. 
The previous equations may be recast as
\begin{equation}
{\dot A}^2 = -k - A^2 \left(\frac{1}{\ell^2} -  
\frac{16 \pi^2 \rho^2}{(d-1)^2\,L_p^{2\,(3-d)}} \right);
\label{9}
\end{equation}
\begin{equation}
\dot \rho = - (d-1)\, (\rho+p)\, \frac{\dot A }{ A}.
\label{energycons}
\end{equation}
Now, we choose a particular equation of state, in the form $p
= p(\rho)$, so as to integrate the conservation
equation (\ref{energycons}) 
\begin{equation}
\ln(A) = -{1\over(d-1)} \int {d\rho\over\rho+p(\rho)}.
\end{equation}
This relationship may be formally inverted to obtain $\rho$
as a function of the brane  ``radius'', $\rho = \rho(A)$. In these 
terms, Eq. (\ref{9}) becomes
\begin{equation}
{\dot A}^2 = -V(A);
\qquad
V(A) = k + A^2 \left(\frac{1}{\ell^2} -  
\frac{16 \pi^2 \rho^2}{(d-1)^2\,L_p^{2\,(3-d)}} \right).
\end{equation}
This single dynamical equation completely determines the expansion of the 
brane.

It is important to stress that when the matter fields become excited
the Hubble rate $\dot{A}/A$ has an extremum, so it is possible to 
expand the dynamical equation around this particular ``radius'' denoted 
by $A_0$. Generically we would have
\begin{equation}
V(A)
=  V(A_0) + V'(A_0)(A-A_0)
 + \frac{1}{2}
V''(A_0)(A-A_0)^2 + {\cal O} ((A-A_0)^3),
\end{equation}
where a prime denotes derivative with respect to $A$.
To compute the various derivatives, we rewrite
the conservation equation as
\begin{equation}
[\rho(A) A ]' = -[ (d-2)\,\rho+(d-1)\, p].
\end{equation}
Differentiating once more we obtain
\begin{equation}
[\rho(A) \,A ]''= \frac{(d-1)\, (\rho + p)}{A} \, [(d-2) + (d-1) v_s^2], 
\end{equation}
where 
\begin{equation}
v_s^2(\rho) \equiv
\left.{\partial p\over\partial\rho}\right|_{\rho}
\end{equation}
is the speed of sound on the brane.
It is easily seen that that the first derivative of the potential
\begin{equation}
V'(A) = \frac{2\,A}{\ell^2} + \frac{32\, \pi^2 \,
\rho\, A\, [(d-2)\, \rho + (d-1)\, p]}{(d-1)^2\, L_p^{2(3-d)}} 
\end{equation}
vanishes if
\begin{equation}
\rho_0 =  \frac{L_p^{(3-d)}}{4\pi} \frac{(d-1)}{A_0}\,
\left(k+\frac{A_0^2}{\ell^2}  \right)^{1/2},
\label{n1}
\end{equation}
and
\begin{equation}
p_0 =  - \frac{L_p^{(3-d)}}{4\pi} \left\{
\frac{(d-2)}{A_0}\left(k+\frac{A_0^2}{\ell^2}\right)^{1/2} +
\frac{A_0/\ell^2}{\sqrt{k + A_0^2/\ell^2}} \right\}.
\label{n2}
\end{equation}
Furthermore,
\begin{equation}
V''(A) = \frac{2}{\ell^2} -\frac{32\, \pi^2 \{
[(d-2)\, \rho + (d-1)\, p]^2 + (d-1) \,\rho\, (\rho+p) 
[(d-2) + (d-1) \,v_s^2] \}}{(d-1)^2\, L_p^{2(3-d)}}
\end{equation}
becomes
\begin{equation}
V''(A_0) = \frac{2}{\ell^2} -  \frac{ 2\,A_0^2/\ell^4}{k + A_0^2/\ell^2}
-  \frac{2\, k}{A_0^2} [(d-2) + (d-1)\, v_{s_0}^2].
\label{oo}
\end{equation}
The square of the expansion velocity is, at this order of approximation,
\begin{equation}
{\dot A}^2 = -\frac{1}{2}V''(A_0) (A-A_0)^2 + {\cal O} ((A-A_0)^3).
\end{equation}
Thus, the equation of motion for the brane requires $V''(A_0)\leq 0$. 
This condition can be re-written in terms of the variable 
$x\equiv A_0^2/\ell^2$ as, (here for  $k=-1,+1$) 
\begin{equation}
\frac{k\,[x - (k+x)\,(d-2)]}{(d-1)} \leq k\,(k+x) \,v_s^2.
\label{oi}
\end{equation}
If we now restrict the speed of sound to lie in the standard 
range: $v_s \in (0,1]$, a glance at Eq. (\ref{oi}) shows that
if $k=-1$, then
\begin{equation}
(3-d)x > (2-d).
\end{equation}  
For $d=2$ this implies that $x$ should be positive, while for $d=3$ the
inequality is trivially satisfied. For $d>3$ we get the following bound
$1<x\leq (d-2)/(d-3)$ -- in $AdS_5$ this is the statement that
$1<A_0^2/\ell^2\leq 2$. 
Note that we have assumed throughout that $x>1$.
Indeed, from Eq. (\ref{metric}) we see that if $x<1$ the brane is 
localized in time and not in the bulk. 
It is noticeable that for $k=-1$ the brane has an
effective cosmological 
constant which is less than zero; if $\dot A_0$ is a maximum,
the system is not able to thermalize to a final state with no cosmological 
constant on the brane. 
Hence, there is no consistent solution minimizing the value
of $\dot A_0$. On the other hand, if $k=1$ (de Sitter instanton) there is 
no constraint on $x$ (in this case, 
Eq. (\ref{oi}) just implies that $x$ is positive), 
and the brane could develop a well--behaved cosmology. 
Finally, for the case $k=0$, there is no bound from the above
considerations since this is a static case.

\subsection{Constraints from the Weyl Anomaly}

Now we will concentrate on the case of $d=4$ and $k=+1$. 
With this in mind, Eq. (12) can be re-written as 
\begin{equation}
\dot{A}^2=-1-\frac{A^2}{\ell^2}\left(
1-\frac{T_{eff}^2}{T_c^2}\right),
\label{carloz}
\end{equation}
where $T_{eff}=T_c+\tilde{\rho}$. 
It should be stressed that an extra term, proportional to $\tilde{\rho}^2$,  
appears in the r.h.s. of Eq. (\ref{carloz}) 
when comparing to the standard Friedmann-Robertson-Walker cosmology 
(a fact already known) \cite{cosmology}. To match the known observations 
of the expanding universe, the latter has to play a  negligible role at 
least back to the time of electron-positron annihilation and 
primordial nucleosynthesis. At this stage we should point out 
that when dealing with compactified extra dimensions, 
one has to stabilize the value of the radion-field (which determines the
size of the extra dimension) at the beginning of 
nucleosynthesis so as not to get into conflict with observations
\cite{radion}. Throughout this paper, however, 
the radion is set to the minimum of its potential. Some constraints on 
the equation of state for cosmology with compactified extra dimensions 
were recently considered in \cite{finland}.

All in all, one expects that the universe  evolves in a similar 
fashion even in the presence of branes at 
temperatures lower than  $T\sim 10^{12}K$ (more on this below).
Nevertheless, there could be a significant departure from the usual scenario at 
very high energy scales -- i.e. at the beginning of the universe -- 
since the expansion 
rate could be dominated by $\tilde{\rho}^2$.     
To find the possible values of $\tilde{\rho}$ 
(at very high energy scales) we
recall that the Weyl anomaly sets the effective tension 
of the brane to be \cite{hhr}
\begin{equation}
T_{eff}=\frac{3\left(1+A_0^2/\ell^2\right)}{4\pi L_pA_0}.
\label{wa}
\end{equation}
The ratio between the energy density of the matter fields and the vacuum
energy at the minimum classical radius of the brane equals
\begin{equation}
\frac{\tilde{\rho}(A_0)}{T_c}=\frac{\ell}{A_0}\left(
1+A_0^2/\ell^2\right)^{1/2}-1.
\end{equation} 
Expanding the square--root with the assumption $A_0\ll \ell$ we see 
that\footnote{Note that if one abandon the idea that the solution come with some string inspired mechanism the $AdS$ radius is not constrained to be of order of Planck length.}
\begin{equation}
\frac{\tilde{\rho}(A_0)}{T_c}\sim \frac{\ell}{A_0}\gg 1.
\end{equation}
This shows that the $\tilde{\rho}^2$-term dominates the early expansion of
the brane world. On the other hand, one can immediately show that 
the Weyl anomaly does not set any constraint on the ratio $A_0/\ell$. 

\section{Present Epoch} 

\subsection{Energy Conditions on the Brane}

The energy conditions, encoded in the evolution of the 
expansion scalar governed by Raychaudhuri's equation \cite{ec}, are 
designed to side-step, as much as possible, the need to pin down a 
particular equation of state. They 
provide simple and robust bounds on the behavior of various linear 
combinations of the components of the stress-energy tensor.  The 
refinement of the energy conditions 
paralleled the development of powerful mathematical theorems, 
such us singularity theorems (guaranteeing, under certain 
circumstances, gravitational 
collapse), the proof of the zeroth law of black hole 
thermodynamics (the constancy of the surface
gravity over the event horizon), 
limits on the extents to which light cones can ``tip over'' in 
strong gravitational fields (superluminal censorship), the cosmic 
censorship conjecture (singularities cannot be unshielded, they 
always remain hidden by event horizons), etc. \cite{ec}. In particular, 
the classical singularity theorem relevant to proving the existence of 
the big-bang singularity relies on the strong energy condition (SEC). 
It is somewhat disturbing to realize that current observations seem to 
indicate that the SEC is violated -- though weakly -- 
somewhere between the epoch of galaxy 
formation and the present time, in an epoch where the cosmological 
temperature never exceeds 60 Kelvin \cite{visser}. 
It is therefore worthwhile to test whether the brane world 
cosmology relaxes or increases the bounds on the violation of the energy 
conditions.
In this section we shall generalize the analysis by Visser 
on Friedmann--Robertson--Walker (FRW) cosmologies \cite{visser}.   

Let us start by setting some basic nomenclature. The weak energy 
condition (WEC) is the assertion that for any timelike vector $\xi^A$, 
$\tilde{T}_{AB} \, \xi^A \xi^B \geq 0$.  
The null energy condition (NEC) is satisfied if and only if,   
$\tilde{T}_{AB} \, \zeta^A \zeta^B \geq 0$ for any null vector $\zeta^A$.
The strong energy condition (SEC) holds if and only if
$(\tilde{T}_{AB} - \frac{1}{2}\, \tilde{T}\, \gamma_{AB}) \geq 0$. 
Finally, the dominant energy condition (DEC) basically says that the locally 
measured energy density is always positive, and that the energy flux is 
timelike or null, that is $\tilde{T}_{AB}\, \xi^A \xi^B \geq 0$, and
$\tilde{T}_{AB} \,\xi^A$ is not spacelike (for an introduction to this
subject, see \cite{ec}). 
These conditions can be rephrased 
in terms of the energy density and the principal pressures as follows,
\begin{itemize}
\item WEC: $\rho>0$, and $\forall j$, $\rho+p_j \geq 0$ 
\item NEC: $\forall j$, $\rho+p_j \geq 0$ 
\item SEC: $\forall j$, $\,\rho+ p_j \geq 0$ and $\rho + \sum_j \, 
p_j \geq 0$
\item DEC: $\rho \geq 0$, and $\forall j$, $p_j \in [-\rho, \rho\,\!]$.
\end{itemize}
where $j=1,\ldots, d-1$.
With these expressions in hand, 
one can easily verify the equivalence between FRW and braneworld cosmology
with respect to WEC, NEC and DEC.
One can also check that 
these energy conditions are not in conflict with the present 
data.\footnote{It should be noted, 
that the energy conditions as defined above refer to the brane world and not 
to the entire spacetime. The extension of these definitions to the whole 
spacetime leads to NEC violation in compactified RS scenarios \cite{007}.}
In the next section we will consider the SEC.

\subsection{What is the Brane World Made of?}

The most direct observational evidence for the expansion of the universe 
comes from the redshift of spectral lines of distance galaxies. 
When we look into the sky and see some object, the look-back 
time $\tilde{\tau}$ to that object is defined as the modulus of the difference 
between $\tau_0$ (the age of the universe now) and $\tau$ the age of the 
universe when the light that we are receiving was emitted. If 
we know the velocity of expansion of the universe $\dot{A}$, by 
putting a lower bound on $\dot{A}$ we deduce an upper bound 
on look-back time. We warn the reader not to confuse $\tau_0$ with 
the birth-time. Throughout this section 
we use the subscript zero to indicate present epoch. Unfortunately 
both usages are standard.

\begin{figure}[htb]
\begin{center}
\epsfig{file=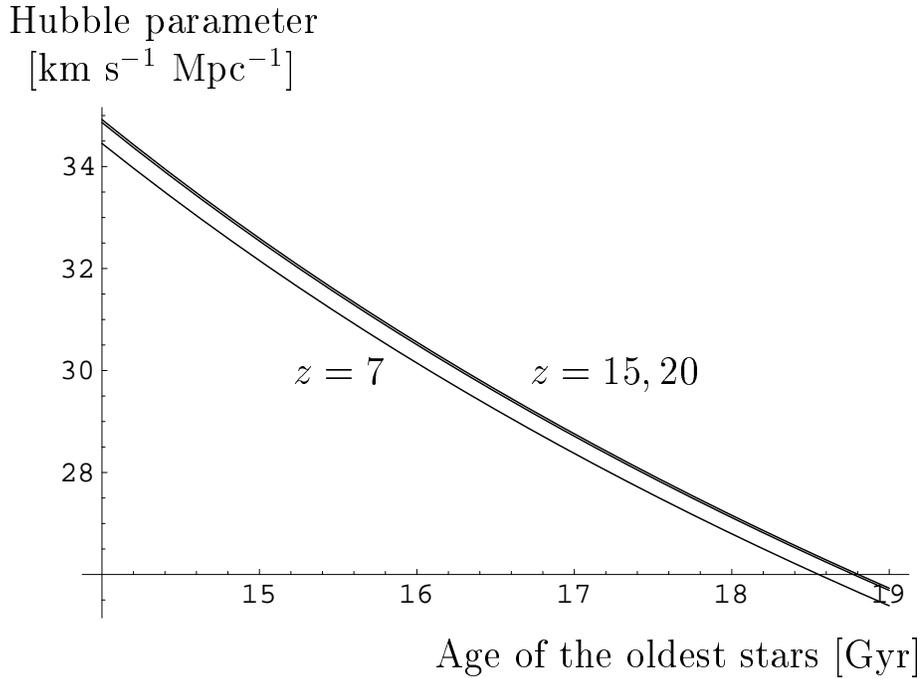,width=14.cm,clip=}
\caption{\baselineskip=12pt {\it Upper bounds on the Hubble parameter $H_0$
as a function of the age of the oldest stars for different redshifts. Note
that the bounds for $z=15, 20$ are almost identical.}}
\end{center}
\label{hubblefigure}
\end{figure}

In particular, from Eqs. (\ref{m1}) and  (\ref{m2}) one can trivially 
check that the SEC is satisfied if and only if
\begin{equation}
(d-3) [k + \dot{A}^2] + (d-2) \frac{A^2}{\ell^2} + \ddot{A} A \leq 0,
\label{ec1}
\end{equation}
and
\begin{equation}
- [k + \dot{A}^2] + \ddot{A} A \leq 0.
\label{ec2}
\end{equation}
Note that the condition (\ref{ec1}) depends on $d$, and for $AdS_3$, 
SEC could be satisfied even with $\ddot{A} > 0$.
Furthermore, contrary to the standard FRW case, SEC shows a $k$ 
dependence on brane-world cosmology. 
We now look for violations of the SEC in our brane cosmology and for that
we concentrate on the condition in Eq. (\ref{ec1}).
We consider a four--dimensional flat brane ($k=0$), 
which according to current experiments is
the most likely to describe the world at the present epoch \cite{tt}.
We further restrict the problem to $A^2\ll\ell^2$.\footnote{Note that to
satisfy SEC $\ddot A$ becomes more negative while increasing $A$ with 
respect to $\ell$.} Let us define the function $f(A) \equiv \dot{A} A$. Using 
Eq. (\ref{ec1}), it is easily seen that 
\begin{equation}
\forall A < A_0, \,\,  f(A) \geq f(A_0).
\label{qwe}
\end{equation}
To set some bounds on the cosmological evolution it is convenient 
to refer the above formulae to the redshift factor $z$. For a photon emitted 
at $\tau_e$, $z$ is defined as   
\begin{equation}
z \equiv \frac{A(\tau_0)}{A(\tau_e)} -1.
\end{equation}
Then after integrating  Eq. (\ref{qwe}) we see that for a flat brane $(k=0)$
the SEC gives the following bound on the Hubble constant:
\begin{equation}
H_0 \leq 
\frac{2z + z^2}{2\,\tilde{\tau}(z) \,(1+z)^2}.
\label{yui}
\end{equation}           

Without belaboring the subject, it has been known for some time 
now \cite{P} that the age of the oldest stars is $16 \pm 2$ Gyr.
The best guess for the redshift at formation of these candles is 
$z_f \approx 15$. Using these values in Eq. (\ref{yui}) yields 
$H_0 \approx 30$ km s$^{-1}$ Mpc$^{-1}$.
Even pulling $z_f$ into $z_f=7$ or $z_f=20$ gives lidicous 
bounds, see Fig. 2. 
Therefore, brane-world cosmology cannot be compatible with 
stellar evolution and the SEC. 
It is easily seen from Eq. (\ref{ec1}) that for $k=-1,+1$ the acceleration 
$\ddot{A}$ has to be more negative and hence that the bounds on $H_0$ are
even more stringent than for the case discussed above.

\section{Final Remarks}

The fate of the universe is still uncertain \cite{tt}. Moreover, 
the possible existence of extra dimensions further complicate the picture.
In this article, we traced a possible evolution of the brane world from 
the very early beginning to the formation of galaxies a few billions 
years later, without enforcing any particular equation of state.
On the one hand, we discussed high energy scales, related to 
the early universe. We see that if the speed of sound is taken  
to lie in the standard range, a de Sitter brane world could develop  
a consistent cosmological scenario, whereas
a similar bound may lead to unusual regimes for $AdS$ domain walls.
This constraint does not depend on the choice of equation of state.
In addition, we find that if $k=1$ the Weyl anomaly increases the effective
tension on the brane in such a way that the matter density $\tilde{\rho}^2$ 
plays a paramount role in the early universe cosmology.  
On the other hand, we have shown that reasonable values of the Hubble 
parameter imply that the strong energy condition 
must be violated sometime between the epoch of 
galaxy formation and present. Consequently, fixing the age of the universe 
does not just imply tuning an equation of state. To overwhelm the 
gravitational effects of the normal matter, we will inescapably need large 
quantities of matter that violates the strong energy condition, or
so--called {\it abnormal} matter. Faced with this fact, it would 
perhaps be interesting to analyze whether any frozen brane-bulk 
interaction could improve the situation. The difficulty with this possibility 
will be maintaining some rather peculiar physics engendered by 
strong energy condition violations.

\bigskip

\subsection*{Acknowledgements} 
We would like to thank Carlos Nu\~nez and Lisa Randall for useful discussions. 
The work of LA was supported by CONICET
and that of KO by the Danish Natural Science Research Council.

\end{document}